\documentclass[graybox]{svmult}

\usepackage{mathptmx}       
\usepackage{helvet}         
\usepackage{courier}        
\usepackage{type1cm}        
\usepackage{makeidx}         
\usepackage{graphicx}        
\usepackage{multicol}        
\usepackage[bottom]{footmisc}

\makeindex             

\begin{document}

\title*{Successful Asteroseismology for a Better Characterization of the Exoplanet HAT-P-7b}
\titlerunning{TDC asteroseismology for characterisation of HAT-P-7b}
\authorrunning{M. Oshagh et al.}
\author{
M.~Oshagh, A.~Grigahc\`ene, O.~Benomar, M.-A.~Dupret, M.~J.~P.~F.~G.~Monteiro, R.~Scuflaire and  N.~C.~Santos}
\institute{
M.~Oshagh, A.~Grigahc\`ene, M.~J.~P.~F.~G.~Monteiro and N.~C.~Santos \at Centro de Astrof\'{\i}sica da Universidade do Porto, Rua das Estrelas, 4150-762 Porto, Portugal \hspace{1cm} \email{moshagh@astro.up.pt}
\and  M.~Oshagh, M.~J.~P.~F.~G.~Monteiro and N.~C.~Santos \at Departamento de F\'{\i}sica e Astronomia, Faculdade de Ci\^encias da Universidade do Porto, Porto, Portugal
\and O.~Benomar \at Sydney Institute for Astronomy (SIfA), School of Physics, University of Sydney, NSW 2006,
Australia
\and M.-A.~Dupret and R.~Scuflaire \at Institut d'Astrophysique et de G\'eophysique, Universit\'e de Li\`ege, All\'ee du 6 Ao{\^u}t 17-B 4000 Li\`ege, Belgium
}


%
%
\maketitle

\abstract{It is well known that asteroseismology is the unique technique permitting the study of the internal structure of
pulsating stars using their pulsational frequencies, which is per se very important. It acquires an additional
value when the star turns out to be a planet host. In this case, the astroseismic study output may be a very
important input for the study of the planetary system.
With this in mind, we use the large time-span of the {\emph{Kepler}} public data obtained for the star system HAT-P-7,
first to perform an asteroseismic study of the pulsating star using Time-Dependent-Convection (TDC) models.
Secondly, we make a revision of the planet properties in the light of the astroseismic study.}

\abstract*{It is well known that asteroseismology is the unique technique permitting the study of the internal structure of
pulsating stars using their pulsational frequencies, which is per se very important. It acquires an additional
value when the star turns out to be a planet host. In this case, the astroseismic study output may be a very
important input for the study of the planetary system.
With this in mind, we use the large time-span of the {\emph{Kepler}} public data obtained for the star system HAT-P-7,
first to perform an asteroseismic study of the pulsating star using Time-Dependent-Convection models.
Secondly, we make a revision of the planet properties in the light of the astroseismic study.}

\section{Introduction}
\label{sec:1}
{\emph{Kepler}} observations are the perfect example of synergy between two separate fields in Astrophysics. This space-mission initially designed to characterize extra-solar planets is supplying the asteroseismology community with unprecedently precise data. The high photometric accuracy reached by {\emph{Kepler}} as required to detect planet transits \cite{2010ApJ...713L..79K} is actually harnessed to undertake asteroseismic studies of great interest in stellar Physics. This implies a better characterization of pulsating stars that harbor planetary systems and consequently tighter constraints on the properties of the latter.

Among the 150,000 stars in the {\emph{Kepler}} field of view a few of them have been identified as planet hosting stars (TrES-2 \cite{2006ApJ...651L..61O}, HAT-P-7 \cite{2008ApJ...680.1450P}, and HAT-P-11 \cite{2010ApJ...710.1724B}) previously to the launch of the satellite. Christensen-Dalsgaard et al. 2010 gave an initial asteroseismic analysis of {\emph{Kepler}} data for these stars \cite{2010ApJ...713L.164C}.

Here we give the preliminary results of our seismic study of HAT-P-7 using the {\emph{Kepler}} public data (Quarters: Q0-Q2). After a precise and careful cleaning of the light-curves from the transits, we obtain the pulsation frequencies. These frequencies are used as input for a TDC modeling of HAT-P-7 oscillations. In turn, the obtained results are used to characterize the orbiting planet.
\section{Kepler data processing and frequency analysis}
\begin{figure}[t]
\centering
\begin{tabular}{cc}
\includegraphics[width=60mm]{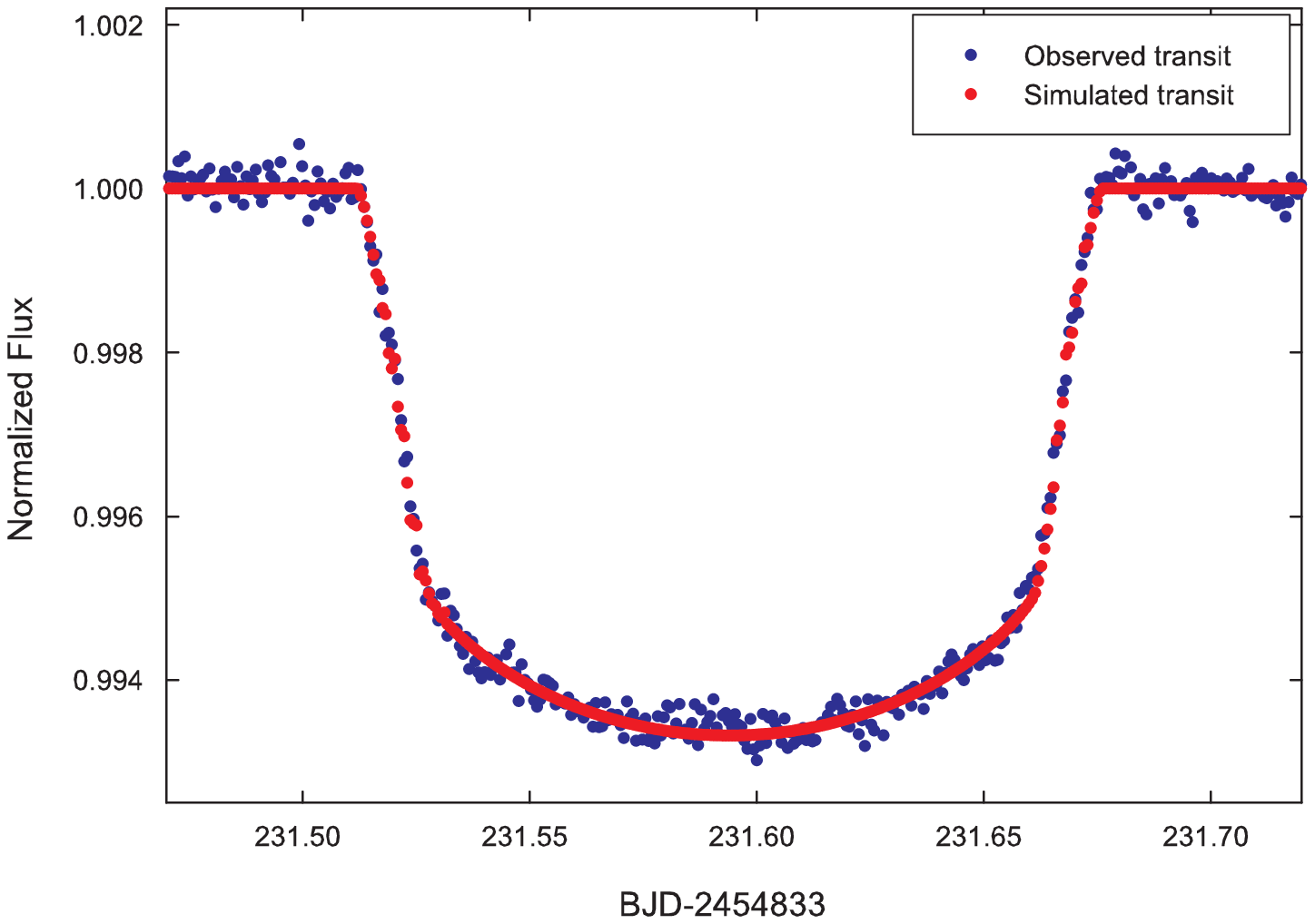} &
\includegraphics[width=60mm]{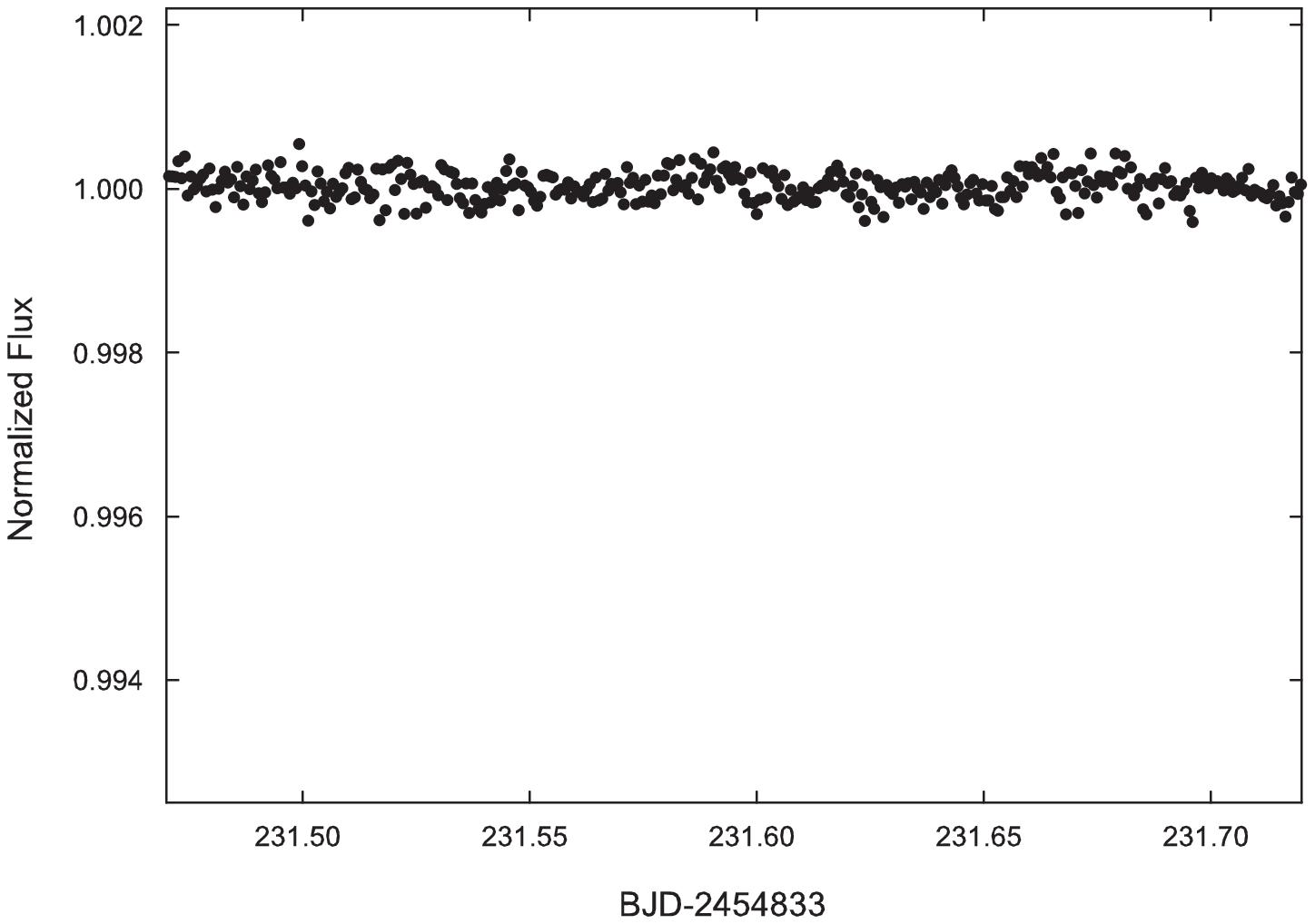}\\
\end{tabular}
\caption{Left: a zoom-in of one of the transits observed in HAT-P-7 light-curve obtained by Kepler. Original observed (blue) and synthetized transit (red) points are shown. See text for more details.
Right: The same light-curve after removing the transit.}
\label{fig1}       
\end{figure}
In this study, we have used public short cadence data gathered by {\emph{Kepler}} for HAT-P-7. The light-curve spans over 144 days, corresponding to the quarters Q0-Q2 \footnote{The data are accessible through the Multimission Archive at STScI at http://archive.stsci.edu/.}.
An efffective astroseismic exploitation of the light curve needs a careful cleaning of the 59 observed transits without loosing the structures inside the transits that may be caused by the star pulsations.
Mandel \& Agol 2002 \cite{2002ApJ...580L.171M} presented a model to produce synthetic transits in photometric light curves. Their method takes into account the stellar and planetary parameters (e.g. star and planet radii, semimajor axis, orbital inclination, stellar limb darkening coefficients) to produce the synthetic transits. It is usually used to constrain the planet parameters.

For our purpose of cleaning HAT-P-7 light curve, we used this method to produce synthetic transits corresponding to the ones observed in HAT-P-7 light curve. We used stellar and planetary parameters reported in \cite{2010ApJ...713L.145W}. We then subtracted the obtained synthetic transits from the observed ones. We ended up with the cleaned light-curve  to be used in the frequency analysis. Fig.~\ref{fig1} gives an example of observed and processed data.

The power spectrum of HAT-P-7, shown in the left panel of Fig.~\ref{Fig2}, has been fitted within a Bayesian framework using a maximum a posteriori approach \cite{2011Benomar}. Here, the main formulated assumption is that frequencies vary smoothly as a function of the radial order $n$. This allows us to improve the robustness and reliability of the extracted frequencies significantly, compared to a classical maximum likelihood approach \cite{2010Appourchaux}.

Using the obtained frequencies and the atmospheric parameters of HAT-P-7 given by \cite{2008ApJ...680.1450P}, we search the best model in a grid of theoretical models. We follow the same methodology as in Grigahc\`ene et al. (these proceedings).

\section{Results}
\begin{figure}[t]
\centering
\begin{tabular}{cc}
\includegraphics[angle=90, width=62mm]{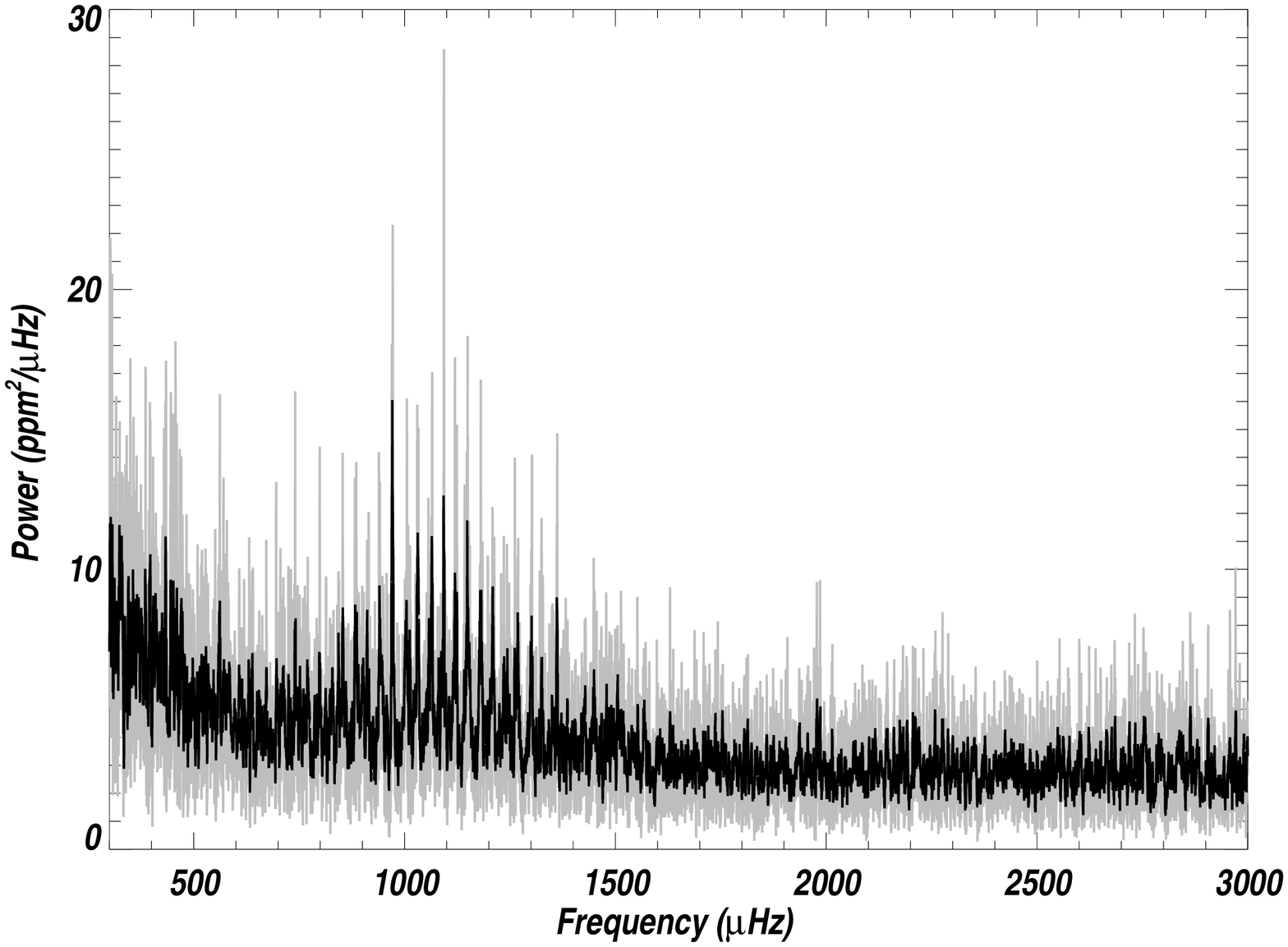} &
\includegraphics[width=63mm]{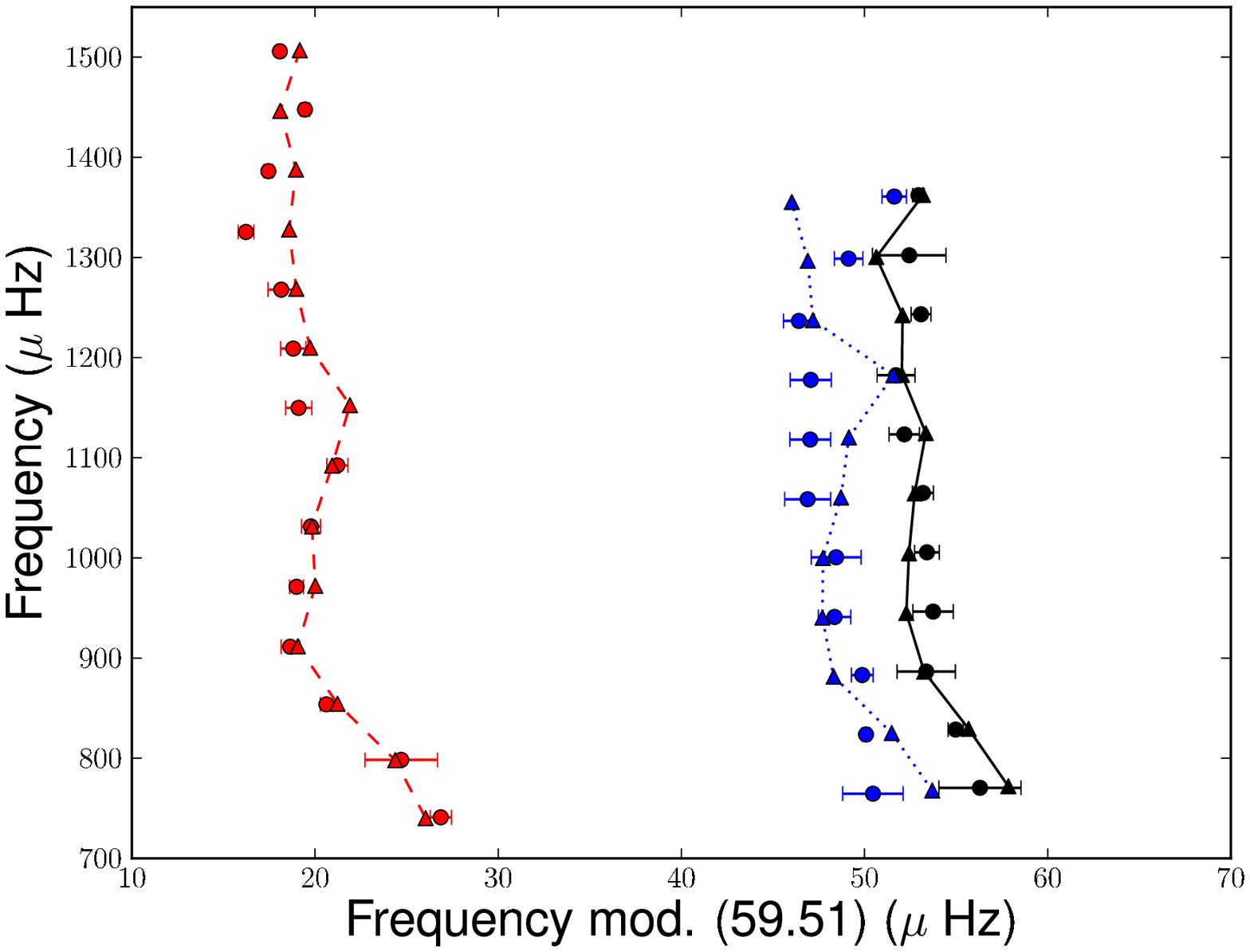} \\
\end{tabular}
\caption{Left: Power spectrum of HAT-P-7.
 Right: \'Echelle diagram for frequencies of degree $\ell$ = 0 - 2 in HAT-P-7, where a large separation of 59.51 $\mu$Hz is used. The filled circles with error bars give the observed frequencies while the filled triangles give the frequencies for the best TDC model. The colors indicate the spherical degree $\ell$: black ($\ell$=0), blue ($\ell$=2) and red ($\ell$=1).}
\label{Fig2}       
\end{figure}

The right panel of Fig.~\ref{Fig2} gives the \'echelle diagram of HAT-P-7. The diagram is built using a large separation of 59.51 $\mu$Hz for frequencies of degree $\ell$ = 0 - 2. The observed (filled circles) and theoretical frequencies (filled triangles) correspond to three identified spherical degrees illustrated by different colors: black ($\ell$=0), blue ($\ell$=2) and red ($\ell$=1). The agreement between model and observations is very good.

The first 4 rows in Tab.~\ref{tab1} give the global parameters of the star. Our results are given in column 4. They are all within the photometric observational boxes. The obtained mass and radius values lie between the isochrones fit values of \cite{2008ApJ...680.1450P} and the adiabatic astroseismic values of \cite{2010ApJ...713L.164C}. TDC models estimation of the age agrees more with the isochrones fitting \cite{2008ApJ...680.1450P} than the adiabatic calculations of \cite{2010ApJ...713L.164C}.

We use then these results as input to characterize the planet. The results are shown in the bottom part of Tab.~\ref{tab1}. We use for that the methods described in \cite{2008ApJ...680.1450P} for the estimation of the dayside temperature and \cite{2010arXiv1001.2010W} for the other parameters. The error bars associated to our estimations are much smaller compared to the others.

\begin{table}[t]
\begin{center}
{\setlength{\tabcolsep}{12pt}
\caption{Stellar and planetary parameters.}
\label{tab1}       
%
%
\begin{tabular}{l  c  c  c}
          & HATNet project       &  {\it Kepler} data: Q0-Q1            & {\it Kepler} data: Q0-Q2                 \\
\hline\noalign{\smallskip}
M$_{\star}$(M$_{\odot}$)        & 1.47$^a$ $\pm$ 0.08     &   1.52$^b$ $\pm$ 0.036  & 1.4150$^d$ $\pm$ 0.020  \\

R$_{\star}$(R$_{\odot}$)        & 1.84$^a$ $\pm$ 0.23     &  1.991$^b$ $\pm$ 0.018   & 1.9276$^d$ $\pm$ 0.010 \\

T$_{\mathrm{eff}}$(K)           &  6350$^a$ $\pm$ 80      &              --          & 6423$^d$ $\pm$ 50 \\

Age(Gy)                     &   2.2$^a$ $\pm$ 1.0      &  2.14$^b$  $\pm$ 0.26    &  2.2082$^d$ $\pm$ 0.04  \\

\noalign{\smallskip}\hline\noalign{\smallskip}

R$_{\mathrm{P}}$(R$_{\mathrm{J}}$) &   1.363$^a$ $\pm$ 0.195 & 1.50$^c$ $\pm$ 0.02      & 1.4588$^d$ $\pm$ 0.007 \\

M$_{\mathrm{P}}$(M$_{\mathrm{J}}$) &  1.776$^a$ $\pm$ 0.077  &  1.82$^c$ $\pm$ 0.03     & 1.709$^d$ $\pm$ 0.024 \\

T$_{\mathrm{dayside}}$(K)         & 2730$^a$ $\pm$ 150      & 2885$^c$ $\pm$ 100       & 2811$^d$ $\pm$ 22 \\
\noalign{\smallskip}\hline\noalign{\smallskip}
\end{tabular}}
\end{center}
$^a$\cite{2008ApJ...680.1450P}: by fitting isochrones.
$^b$\cite{2010ApJ...713L.164C}: Adiabatic astroseismic study.
$^c$\cite{2010ApJ...713L.145W}.
$^d$This work.
\end{table}

In summary, our physically more robust TDC modeling of the oscillations of HAT-P-7 yields more reliable global parameters for the star. Which gives tighter constraints on the properties of its orbiting planet. The analysis of the remaining data obtained by {\emph{Kepler}} is expected to give more precise frequency list.
We plan to run more models in order to build a denser grid. This will allow us to explore a larger parameter space for a better characterization of the star and hence, its planet.

\begin{acknowledgement}
   Part of this work was supported by \emph{FCT-MCTES-Portugal} and
   \emph{FEDER-EC},
   through grants {\scriptsize PTDC/CTE-AST/098754/2008},
   {\scriptsize SFRH/BPD/41270/2007} and {\scriptsize SFRH/BPD/47611/2008}.
\end{acknowledgement}

\end{document}